\documentclass[preprint,12pt]{elsarticle}

\usepackage{amssymb}
\usepackage{amsmath}
\usepackage{multirow} 
\usepackage[dvipsnames]{xcolor}

\journal{Journal of Magnetism and Magnetic Materials}

\begin{document}

\begin{frontmatter}

\title{Effect of 3\textit{d} Transition Metal Doping (Mn, Fe, Co, Ni) on the Electronic and Magnetic Properties of Pd Alloys at Low Impurity Concentrations: An \textit{Ab initio} Study}

\author{Irina I. Piyanzina $^{a,b*}$, Zarina I. Minnegulova$^{b}$, Regina M. Burganova$^{b}$,  Oleg V. Nedopekin$^{b}$, Igor V. Yanilkin$^{b}$, Vasiliy S. Stolyarov$^{c}$, Amir I. Gumarov$^{b}$  \\ 
i.piyanzina@gmail.com}

\affiliation{organization={Center of Semiconductor Devices and Nanotechnology, Computational Materials Science Laboratory, Yerevan State University, Republic of Armenia},    
addressline={1 Alex Manoogian St.}, 
           city={Yerevan},
            postcode={0025}, 
           country={Republic of Armenia}}

\affiliation{organization={Institute of Physics, Kazan Federal University},
            addressline={16~Kremlyovskaya~str.}, 
            city={Kazan},
            postcode={420008}, 
            country={Russia}}

\affiliation{organization={Center for Advanced Mesoscience and Nanotechnology, Moscow Institute of Physics and Technology},
            addressline={Institutskiy per 9}, 
            city={Dolgoprudniy},
            postcode={141701}, 
            country={Russia}}

\begin{abstract}
The nature of low-impurity ferromagnetism remains a challenging problem in the solid-state community. Despite initial experiments dating back to the mid-20th century, a comprehensive theoretical explanation and reliable \textit{ab initio} evaluations have remained elusive. The present research aims to bridge this gap by refining first-principle calculations by elucidating the magnetic and electronic behavior of Pd$_{1-x}$$M_{x}$ alloys (where $M$ = Mn, Fe, Co, Ni). Our study includes calculations of magnetic properties throughout the range of impurity concentrations, from 1 to 100 atomic percent (at.\%), where we estimate critical concentrations and perform a comparative analysis for the listed alloys. Furthermore, electronic structure was analyzed, including the calculations of atomic, spin, and orbital-resolved states density, and exploration of the spatial formation of magnetic clusters containing ferromagnetic impurities across all concentration ranges.
\end{abstract}

\begin{highlights}
\item The DFT methodology was adapted to predict the properties of low-impurity alloys. 
\item DFT and experimental magnetic moment dependencies on concentration showed good agreement.
\item A sharp increase in magnetization was observed at low concentrations.
\item The PdNi and PdMn alloys demonstrated a critical concentration of 3\,at.\%;
\item The electronic structure of Pd$_{1-x}$Ni$_{x}$ alloy has been thoroughly analyzed. 
\end{highlights}

\begin{graphicalabstract}
\includegraphics[width=\linewidth]{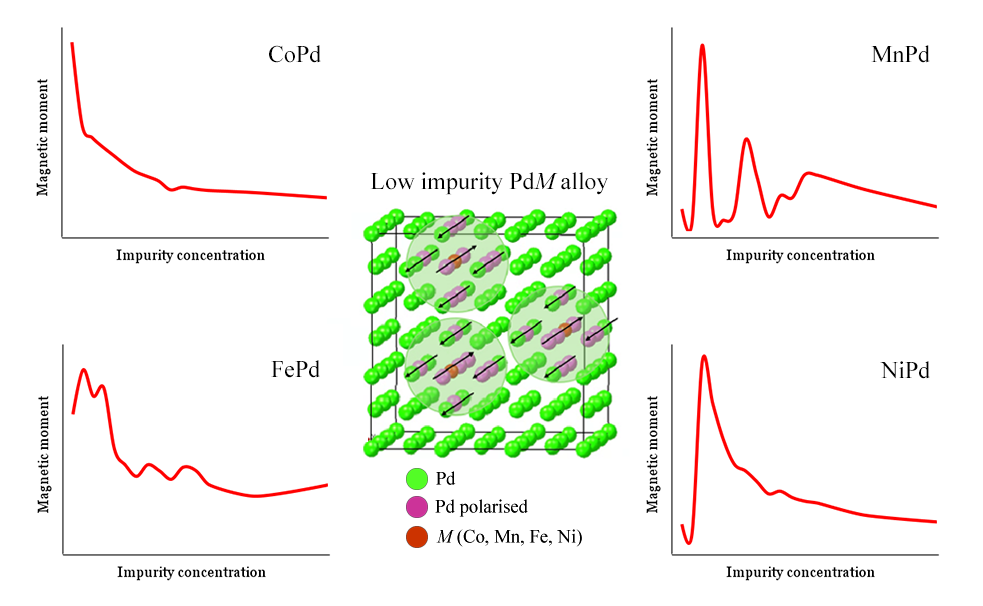}
\end{graphicalabstract}


\end{frontmatter}

\section{Introduction}
\label{intro}

The exploration of Pd$_{1-x}$\textit{M}$_{x}$ alloys (\textit{M} is ferromagnetic impurity) dates back to 1938, with foundational research credited to Fallot~\cite{fallot}. Subsequent in-depth examinations of bulk Pd$_{1-x}$Fe$_{x}$ alloys were conducted by Crangle in 1960~\cite{crangle}. The theoretical background of the arising phenomenon was presented in the fundamental paper of Korenblit and Shender~\cite{Korenblit}. The appearance of impurity ferromagnetism in paramagnetic metals was explained there by the indirect interaction of impurity spins through strongly interconnected electrons in a narrow \textit{d}-band. As a result of this interaction, the spin of the impurity is surrounded by an electron cloud polarized along or against the impurity moment. This was demonstrated, for example, for NiPd and NiPt alloys, where Ni was found to have different signs of magnetic moment~\cite{singh}. In its turn, the overlap of the polarized electron clouds surrounding the impurity leads to an indirect exchange interaction between the impurities. In the ferromagnetic phase, the presence of magnetic moments of the localized spins leads to magnetization of the host Pd matrix near the impurity.

It is well-known that ferromagnetism exists in materials such as iron (Fe), cobalt (Co), or nickel (Ni), and in some rare earths and their alloys. The stability of ferromagnetism in these elements can be explained within the framework of the Stoner criterion~\cite{Stoner}, which determines the threshold at which ferromagnetism arises in the system of free electrons. This criterion can be described by the inequality: 
\begin{equation*}
\begin{gathered}
    D(E_F) \times I > 1,
\end{gathered}
\end{equation*}
where \textit{I} is the coefficient of exchange interaction of itinerant electrons, and \textit{D(E$_F$)} is their density of states at the Fermi level. The criterion gives the ratio of the exchange and kinetic energies of electrons necessary for their ferromagnetic ordering. 
The transition from paramagnetism to ferromagnetism becomes preferable if the inequality is satisfied, as in Fe, Co, Ni, or Mn. Then the system can reduce its energy by pumping a sufficient number of polarized electrons depending on the direction of the spin and their energy, thereby leading to a splitting of the levels and the formation of an exchange interaction or, more simply, to ferromagnetism.
Palladium and platinum are materials that nearly satisfy the Stoner criterion and consequently called nearly magnetic materials. When they are alloyed, they transform from a paramagnetic to a ferromagnetic state.

The first experimental works demonstrated that the addition of iron to palladium causes an anomalously large magnetic moment. In particular, Nieuwenhuis showed in his experiments that ferromagnetism occurs at very low concentrations of impurities (less than 1\,at.$\%$), and the magnetic moment can reach 10 $\mu_B$ per Fe and Co atoms~\cite{nieu}. In addition, a decrease in the magnetic moment is observed with an increase in the impurity concentration. A similar dependence is observed in almost all experimental works for such alloys, but alloys with palladium, as a rule, have a larger magnetic moment than with platinum~\cite{singh}.

On the basis of this, the study of the effect of ferromagnetic impurity on the magnetic properties of binary palladium alloys is of great importance for science and can lead to the creation of new materials with improved properties for various fields of application. Thus, recently, interest in the structures of the "superconductor-ferromagnet-superconductor" type has increased due to the potential use in superconducting spintronics~\cite{ryazanov,arham,bolginov,larkin,ryazanov2,vernik,niedziel,glick,soloviev,usp}. One of the application options for Pd-e alloys is in superconducting magnetic random access memory (MRAM) based on Josephson junctions~\cite{ryazanov,larkin,ryazanov2,soloviev,Golovchanskiy}. In this context, Pd-rich ferromagnetic compositions within the impurity concentration range $0.01 < x < 0.1$ are particularly relevant. These alloys are typically used in the form of thin films, manufactured using techniques such as magnetron sputtering~\cite{arham,bolginov,larkin,ryazanov2,vernik,niedziel,glick,usp,uspenskaya1,uspenskaya2,bolginov2}, molecular beam epitaxy (MBE)~\cite{uspenskaya2,garifullin,ewerlin,esmaeili,esmaeili2,petrov,esmaeili3,mohammed,ya}, and ion beam implantation~\cite{gumarov}.

Recent experimental studies have shown that the addition of ferromagnetic impurities such as Fe to palladium (Pd) can significantly change their magnetic properties~\cite{esmaeili,esmaeili2,petrov,esmaeili3,mohammed,ya,gumarov,gumarov2}. In particular, the addition of Fe and Co can increase the coercive field and saturation magnetization, while adding low concentrations of Ni and Mn to palladium alloys can achieve a low coercive field and low saturation magnetization, making these alloys more suitable for MRAM applications. At the same time, all listed impurities affect palladium in different ways; for example, at high manganese concentrations, antiferromagnetic compounds may form, which can reduce the magnetic properties of the alloy. Thus, choosing the optimal concentration and type of ferromagnetic impurities in palladium alloys for MRAM applications requires a balance between the magnetic properties and the alloy structure.

Limited theoretical and first-principles research exists on the electronic structure of Pd$_{1-x}$\textit{M}$_{x}$ solid solutions~\cite{opahle,drittler,burzo,golovnia,shi,singh}. In particular, these studies did not present predictions and analysis of electronic and magnetic properties for an ultra low concentration region of $x < 0.1$, rendering the findings inapplicable to Pd-rich alloys. The only computational examples of density functional theory (DFT) for low impurity concentrations were realized in our previous work, where we analyzed the concentration and position dependencies for the Fe impurity~\cite{piyanzina_crystals} and comparison with experimental measurements for Co~\cite{piyanzina_surface}.

In this work, we present a comprehensive $\textit{ab initio}$ analysis of the magnetic and electronic properties of Pd$_{1-x}$$M_{x}$ alloys, where  \textit{M} = Mn, Fe, Co, or Ni. Metal dopants are uniformly distributed in the Pd host lattice across various impurity contents $x$, such as $0.01 < x < 1$, and we calculate the mean magnetic moment per impurity atom as a function of $x$. To study the electronic configuration of the arising magnetic states and the spatial distribution of magnetic clusters, we analyze the atomic, spin, and orbital resolved density of states for alloys with different impurity concentrations.

\section{Computational Details}
\label{computations}

Our $\textit{ab initio}$ investigations were based on the DFT~\cite{hohenberg1964,kohn1965} approach within the VASP code~\cite{kresse1996a,kresse1996b,kresse1999} as part of the MedeA\textsuperscript{\textregistered} software of Materials Design~\cite{medea}. The effects of exchange and correlation were taken into account by the generalized gradient approximation (GGA)  as parameterized by Perdew, Burke, and Ernzerhof (PBE)~\cite{perdew1996}. The Kohn--Sham equations were solved, using the set of plane waves (PAW)~\cite{bloechl1994paw}. The cut-off energy was chosen to be equal to 400\,eV. The force tolerance was 0.5\,eV/nm and the energy tolerance for the self-consistency loop was 10$^{-5}$\,eV. The Brillouin zones were sampled using Monkhorst--Pack grids~\cite{monkhorst1976}, including 5 $\times$ 5 $\times$ 5 {$\textbf{k}$-}points for alloys and 7 $\times$ 7 $\times$ 7 for bulk cells of pure components. We performed spin-polarized calculations in all cases, initializing impurity ions to have initial magnetic moments, and Pd atoms to be in the paramagnetic state. The electronic densities of the states were calculated using the linear tetrahedron method~\cite{bloechl2} on the same {\bf k}-point grids with 0.03 integration step.

The structures are described as consisting of a filled face-centered cubic (FCC) host matrix formed by Pd atoms with impurity ions substituting octahedrally coordinated sites only. The unit cell of the alloy was contracted as 3$\times$3$\times$3 bulk unit cells of Pd with a homogeneous distribution of impurities preserving the Pd lattice parameter.

For bulk calculations, the full three-step optimization was performed with increased accuracy at each step, whereas for alloys, because of computational restrictions, the relaxation of atomic coordinates was only performed.

Additionally, the simplified GGA+\textit{U}~\cite{Dudarev} approach was applied to take into account strong correlations between electrons. In the case of PAW potential while considering \textit{f}-electrons as valence states, we used the values of \textit{U}$_d$ for the Pd, Mn, Fe, Co and Ni ions according to Refs.~\cite{aflow,ostlin,hong}, as collected in Table~\ref{tab:bulk}.

\section{Results}
\subsection{Bulk characteristics}
\label{bulk}

In the first stage, we present the results for the bulk components of the alloys under consideration in order to provide a detailed analysis of the resulting electronic and magnetic features. Methodological parameters were also tested, comparing the calculated and experimental lattice constants and magnetic moments for FCC cells with the symmetry of the Fm-3m space group. Since Co can exist in other phases, we present the results for its hexagonal phase as well. As shown in Table\,\ref{tab:bulk}, the calculated lattice parameters agree very well with the available experimental data, supporting the precision of the proposed computational scheme.
Moreover, the calculated lattice parameters for Pd, Ni and Co are consistent with previously published linearized augmented planewave (LAPW) method calculations~\cite{shi}, which reported lattice sizes of 3.89, 3.54 and 3.52\,\AA ~for Pd, Co and Ni, respectively. 
\begin{table}
\centering
   \caption{Used Coulomb parameter U$_d$ in eV~\cite{aflow,ostlin,hong}, optimized lattice constants \textit{a}=\textit{b}=\textit{c} (except of Co hcc) (all in \AA), magnetic moments per ion (in $\mu_B$) of the bulk Pd, Mn, Fe, Co, and Ni comparing to the experimental data where available.}
\begin{tabular}{l|llllll}
Bulk cell & \textit{U$_d$} & a, \AA  & a, exp & $\mu_B$ & $\mu_B$, exp 
\\ \hline
Pd   & 2.0         & 3.898 &   3.876 (100\,K)~\cite{king}     & 0.407 &        
\\
Mn   & 8.0         & 4.216 &        & 4.407 &        
\\
Fe   & 4.6         & 3.715 &        & 2.980 &    2.22~\cite{shull}    
\\
Co fcc  & 5.0         & 3.560 &   3.5446~\cite{ishida}      & 1.987 &   1.61~\cite{Meyers}      
\\
Co hcc  & 5.0         & a=2.507 c=4.07 &         & 2.085 &        
\\
Ni   & 6.0         & 3.447 &        & 0.908 &    0.616~\cite{singh,cadeville}     
\\            
\end{tabular}
\label{tab:bulk}
\end{table}

\begin{figure}
\centering
\begin{minipage}{0.49\textwidth}
 \includegraphics[width=\linewidth]{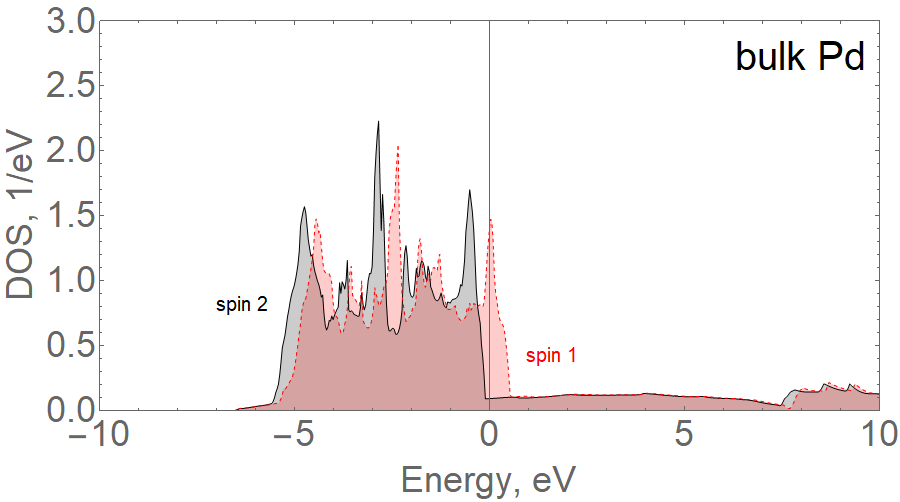} \\ (a)
\end{minipage}\\
\begin{minipage}{0.49\textwidth}
 \includegraphics[width=\linewidth]{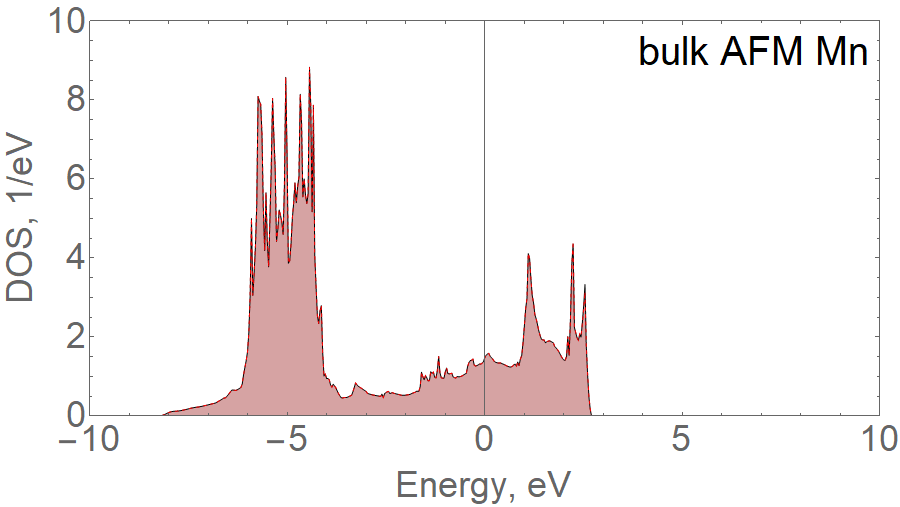} \\ (b)
 \end{minipage}
\begin{minipage}{0.49\textwidth}
 \includegraphics[width=\linewidth]{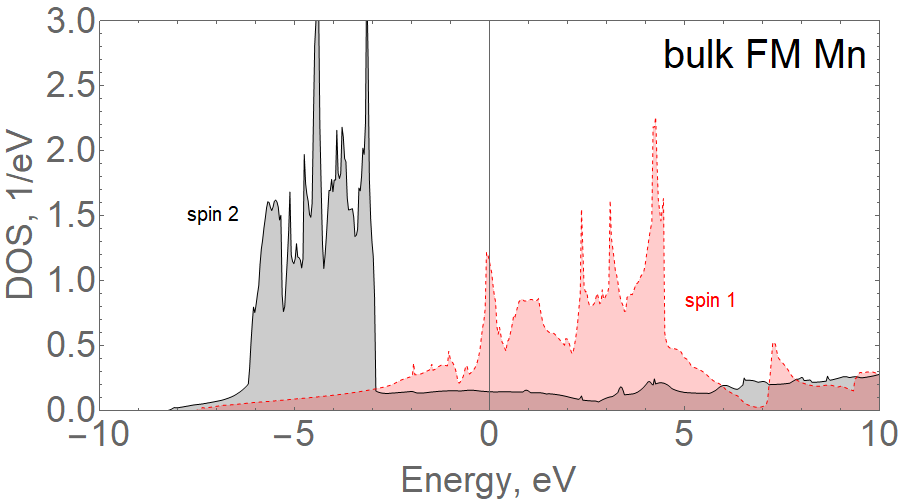} \\ (c) 
\end{minipage}
\begin{minipage}{0.49\textwidth}
 \includegraphics[width=\linewidth]{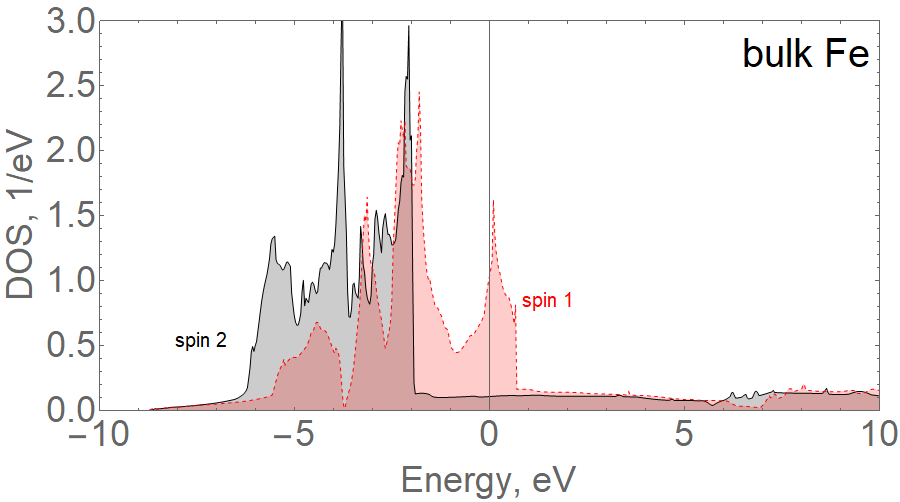} \\ (d)
\end{minipage}
\begin{minipage}{0.49\textwidth}
 \includegraphics[width=\linewidth]{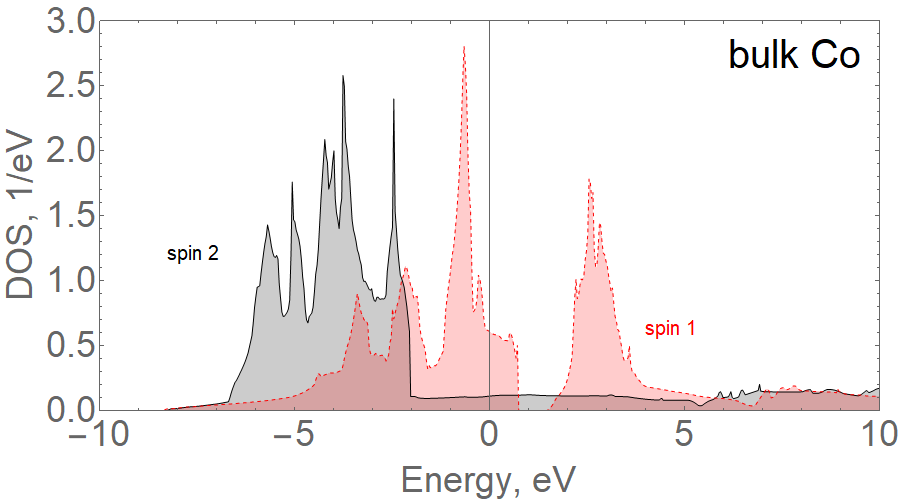} \\ (e) 
\end{minipage}
\begin{minipage}{0.49\textwidth}
\includegraphics[width=\linewidth]{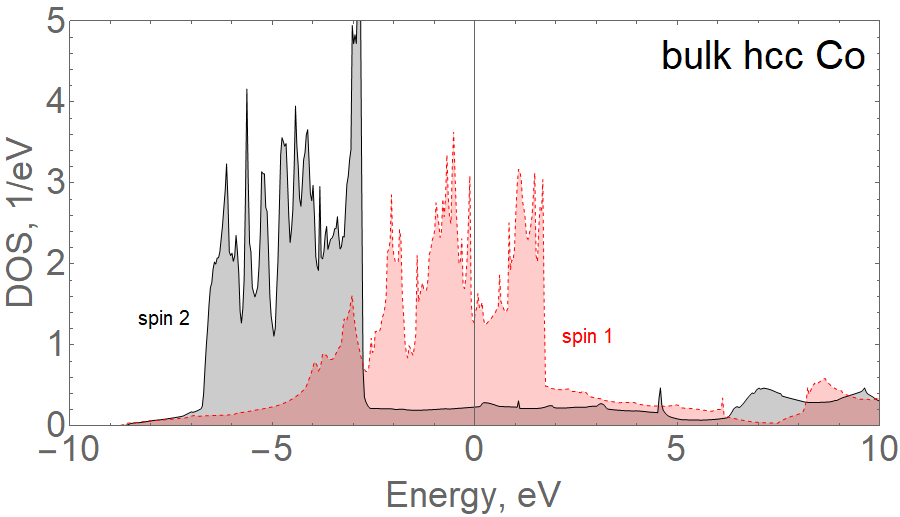} \\ (f) 
\end{minipage}
\begin{minipage}{0.49\textwidth}
 \includegraphics[width=\linewidth]{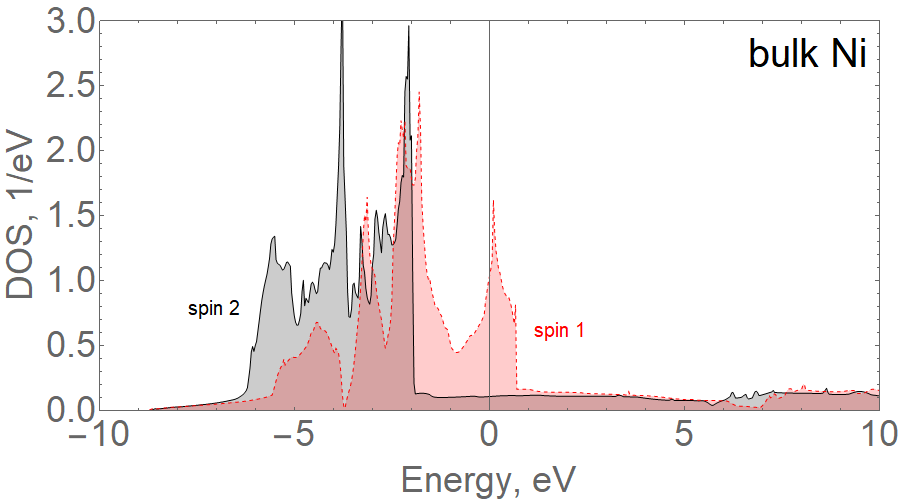} \\ (g)
\end{minipage}
\caption{Calculated density of states for bulk Pd, Mn, Fe, Co, Ni. The red and black lines denote different spin components. Mn in ferromagnetic (FM) and antiferromagnetic (AFM) states is presented. DOS for Co is given for the fcc and hcc phases.}
\label{fig:bulk_dos}
\end{figure}

The calculated spin-resolved density of states (DOS) plots for the considered bulk compounds are presented in Figure\,\ref{fig:bulk_dos}. 
As mentioned in Section\,\ref{intro} palladium (Pd) is a weak magnetic material that satisfies the Stoner criterion. That fact is supported by DOS plot (Fig.\,\ref{fig:bulk_dos}\,a), which shows little spin splitting resulting in 0.407\,$\mu_B$ magnetic moment per atom. This value is in good agreement with previous calculations~\cite{moruzzi1989magnetism, chen1989electronic}, where the maximum magnetic moment in ferromagnetic Pd was found to be 0.35\,$\mu_B$. Moreover, both spin components are located in the same energetic region with a similar distribution.

Manganese is a transition metal with half-filled 3\textit{d} orbitals, it has five unpaired electrons, giving it the highest magnetic moment among the compounds studied of 4.407\,$\mu_B$ (Table\,\ref{tab:bulk}). This half-filled 3\textit{d} subshell results in a high-spin configuration, which promotes antiferromagnetic (AFM) ordering. We perform DOS calculations for the ferromagnetic (FM) and AFM ordering (Fig.\,\ref{fig:bulk_dos}\,b and c), which revealed that Mn has similar energies for both configurations, making them equally probable. In particular, FM ordering shows significant differences in spin occupancy, leading to the largest energy split between the spin components (as seen in Fig.\,\ref{fig:bulk_dos}\,c). 

Iron, cobalt, and nickel have similar spin distributions because of their analogous electronic configurations. The difference in their magnetic moments arises from the number of unpaired electrons: 4 for Fe, 3 for Co, and 2 for Ni. In all three elements, one spin component has more states located above the Fermi level, while the other has more states below, resulting in their metallic ferromagnetic behavior.

\subsection{Concentration dependencies for alloys}
\label{magnetic}

The concentration dependencies for all alloys with impurity levels below 25\,at.$\%$ were initially analyzed, revealing a relatively large magnetic moment per ferromagnetic impurity. A detailed analysis for each case will be presented separately.

\subsubsection{Mn impurity alloy}
\label{Mn_magnetism}

\begin{figure}
\centering
\begin{minipage}{0.49\textwidth}
 \includegraphics[width=\linewidth]{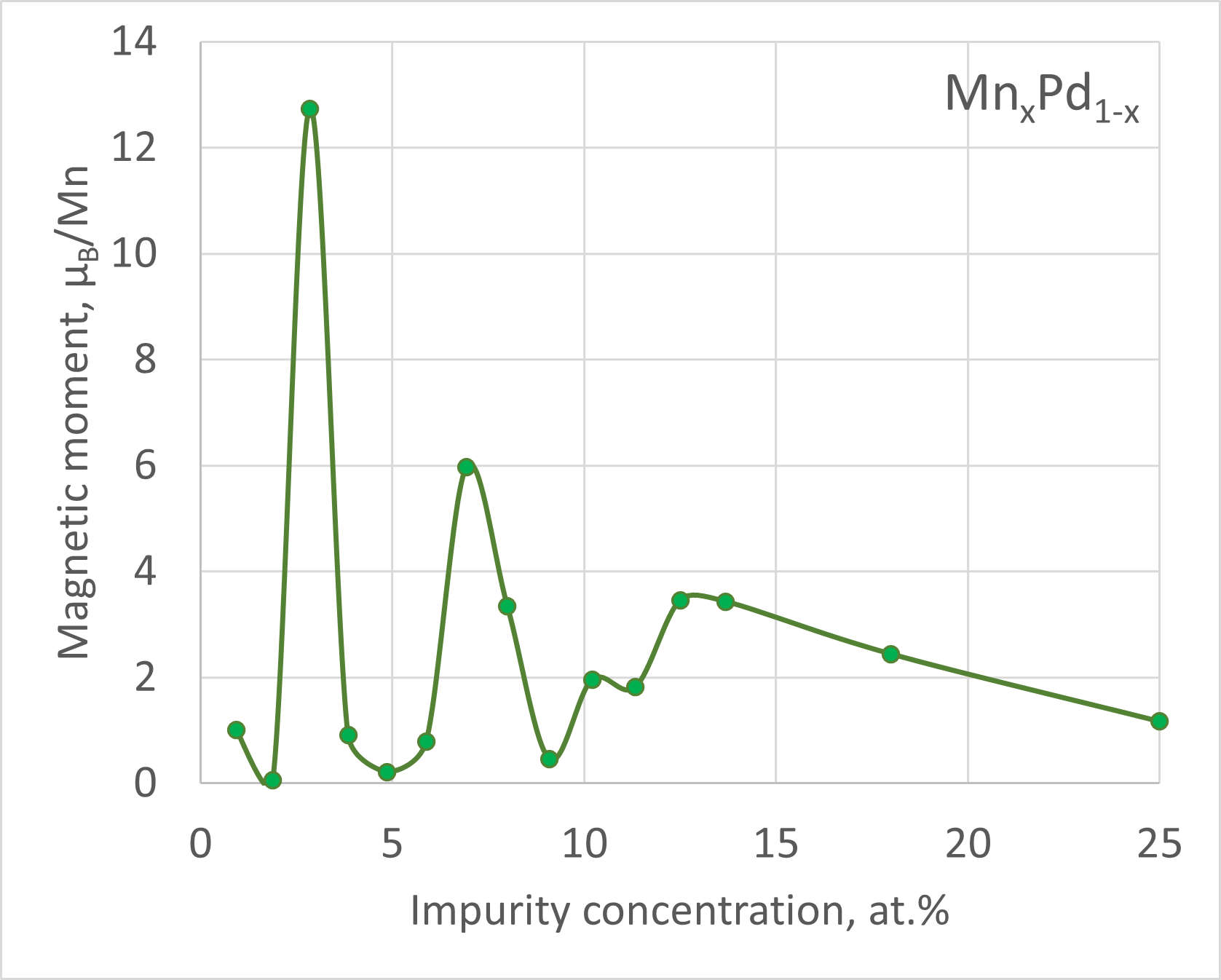} \\(a) 
\end{minipage}
\begin{minipage}{0.49\textwidth}
 \includegraphics[width=\linewidth]{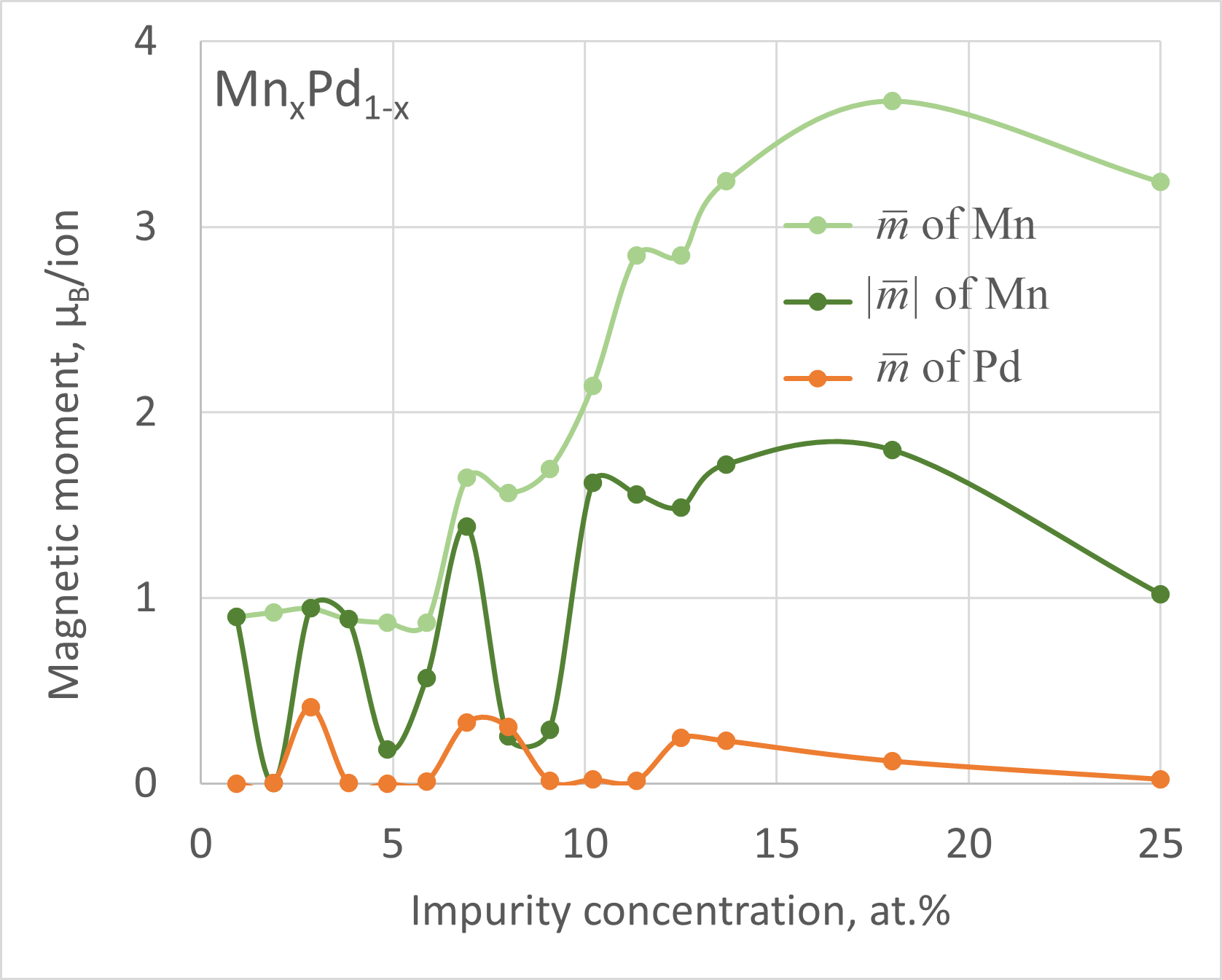} \\(b) 
\end{minipage}
\caption{(a) Calculated magnetic moment (in $\mu_B$) per impurity ion for the Mn$_x$Pd$_{1-x}$ alloy versus Mn concentration (in at.\%) up to 25 at.\%. (b) Calculated mean magnetic moment (in $\mu_B$) of Pd and Mn ions  (in dark green) and calculated mean of the absolute magnetic moments of impurity ions (in light green) versus impurity concentration (in at.\%) up to 25\,at.\%.}
\label{fig:mn}
\end{figure}

The alloy with manganese (Mn) impurities shows alternating maxima and minima in the magnetic moment per Mn ion as a function of impurity concentration (Figure\,\ref{fig:mn}\,a). We associated this peculiar behavior with the antiferromagnetic nature of Mn. To better understand this phenomenon, we also plotted the magnetic moment dependencies of Mn ions, averaged over all impurity ions, in two ways: the mean magnetic moment and the mean of the absolute values of the magnetic moments, as illustrated in Figure\,\ref{fig:mn}\,b. The mean magnetic moment exhibits a sequence of maximums and minimums. The minimums correspond to the compensation effect, i.e. an equal number of positive and negative magnetic moments of Mn ions. In contrast, maximums reflect situations where magnetic moments are uncompensated. In contrast, the mean of the absolute values shows a gradual increase in the Mn magnetic moment without alternation, approaching the bulk Mn value.
 
Furthermore, the mean magnetic moment of the Pd ions follows a similar trend to that of Mn, displaying maximums and minimums (Figure\,\ref{fig:mn}\,b), in line with the antiferromagnetic behavior of Mn. 

\begin{figure}[ht!]
\centering
\begin{minipage}{0.45\textwidth}
\includegraphics[width=\linewidth]{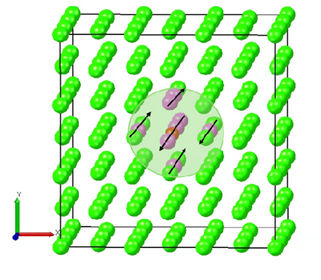} \\(a) 
\end{minipage}
\begin{minipage}{0.45\textwidth}
\includegraphics[width=\linewidth]{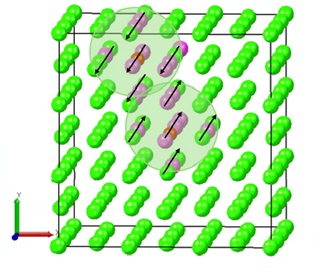} \\(b)
\end{minipage}
\begin{minipage}{0.45\textwidth}
\includegraphics[width=\linewidth]{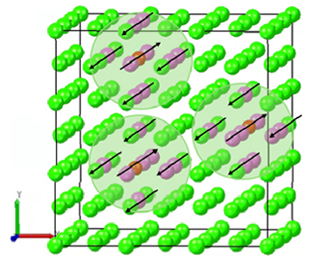} \\(c)
\end{minipage}
\begin{minipage}{0.45\textwidth}
\includegraphics[width=\linewidth]{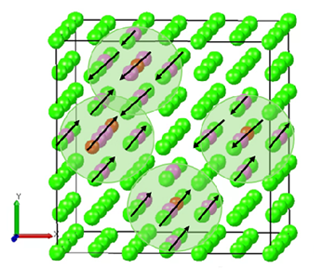} \\(d)
\end{minipage}
\caption{Representation of the magnetic cluster formation in Mn$_x$Pd$_{1-x}$ (0 $<$ x $<$ 0.04) alloy for 1-4 at.\% respectively (a-d). The green spheres represent Pd matrix, the pink spheres denote mostly polarized Pd ions, and the red spheres are Mn ions. The arrows denote the predominant direction of magnetic moments within the clusters.} 
\label{fig:Mn_claster}
\end{figure}

Figure\,\ref{fig:Mn_claster} explains alternating maximums and minimums in plots of Figure\,\ref{fig:mn}\,a,b in therms of the formation of magnetic clusters in Mn$_x$Pd$_{1-x}$ (0$<$x$<$0.04).
At concentrations of 1 and 2\,at.\%, antiferromagnetic order is observed for both manganese and palladium (Figure\,\ref{fig:Mn_claster}\,a,b). At a concentration of 3\,at.\%, which can be considered critical (also in agreement with the plot in Figure\,\ref{fig:mn}\,a), the magnetic moments of Mn and Pd coincide, but align in opposite directions. As a result of this alignment, the overall magnetic moment of the system increases (Figure\,\ref{fig:Mn_claster}\,c). At higher concentrations of Mn, antiferromagnetic order reappears, with pairs of manganese ions compensating for the magnetic moments of each other, and similarly the magnetic moments of pairs of Pd ions become compensated (Figure\,\ref{fig:Mn_claster}\,d).

The comparison with the experimental data for this alloy is ambiguous, as the available data show significant variation. According to Nieuwenhuys' review~\cite{nieu}, a magnetic moment of 7-8\,$\mu_B$ per Mn ion has been reported. Furthermore, some studies have shown that the ferromagnetic ordering disappears below 5 at.\% of Mn~\cite{nieu,zweer}. These findings contradict our calculations, but this discrepancy might be explained by both the different experimental conditions and the complex magnetic nature of Mn, which could exhibit two distinct magnetic phases.

\subsubsection{Fe impurity alloy}
\label{Fe_magnetism}

\begin{figure}
\centering
\begin{minipage}{0.49\textwidth}
 \includegraphics[width=\linewidth]{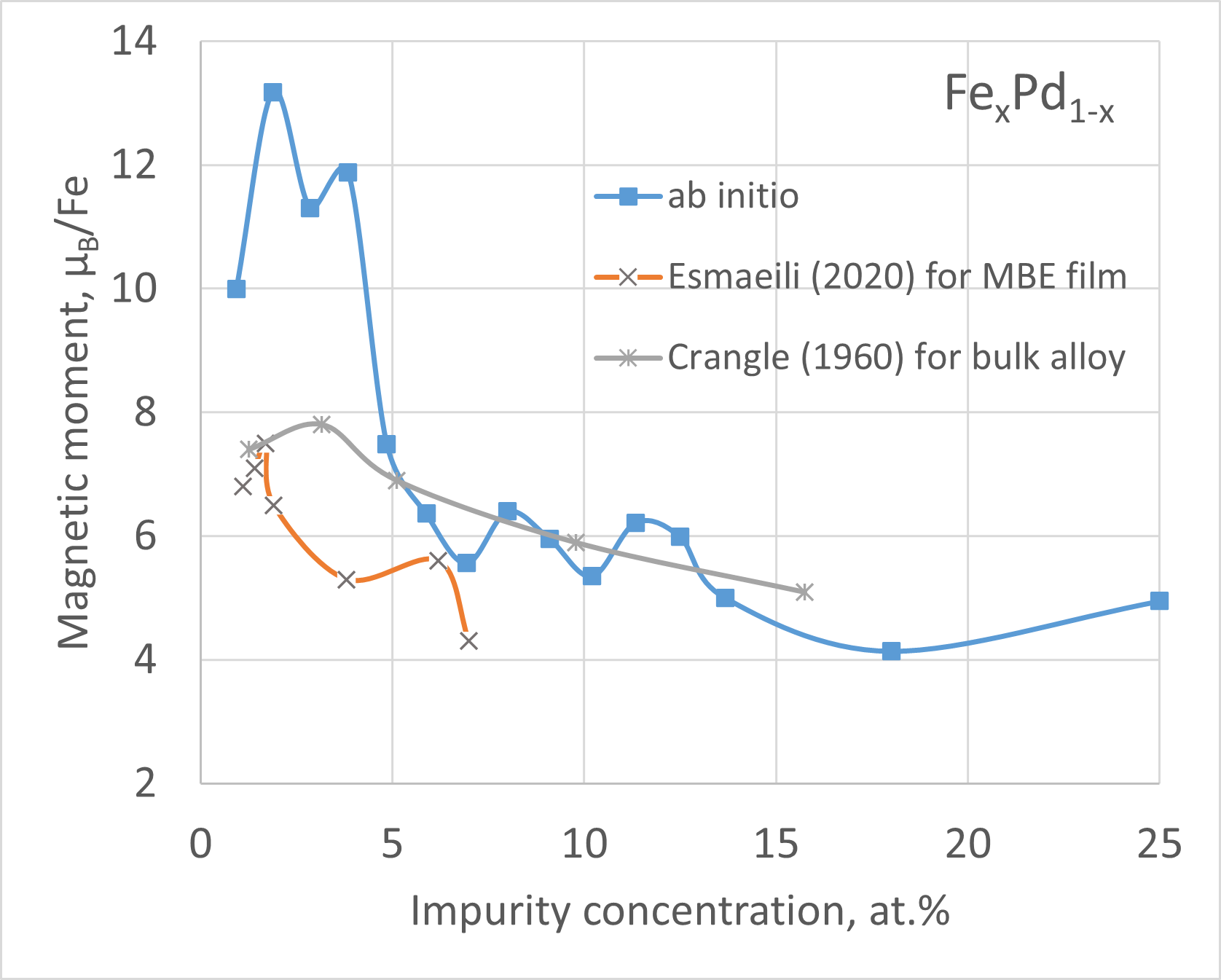} \\(a)
\end{minipage}
\begin{minipage}{0.49\textwidth}
 \includegraphics[width=\linewidth]{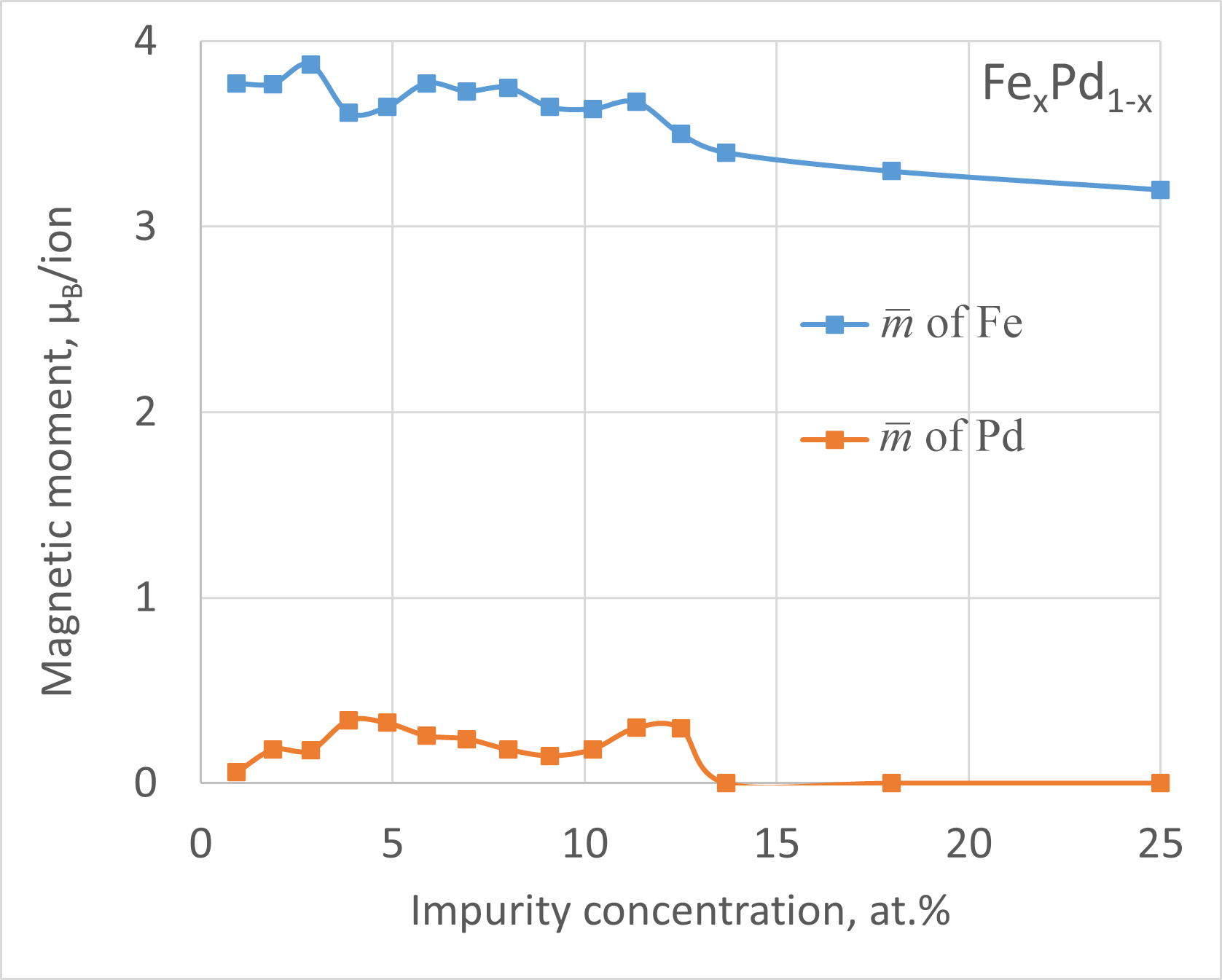} \\(b)
\end{minipage}
\caption{(a) Calculated magnetic moment (in $\mu_B$) per impurity ion for the Fe$_x$Pd$_{1-x}$ alloy versus Fe concentration (in at.\%) up to 25 at.\%. Available experimental data from Refs.~\cite{crangle,esmaeili2} is presented as well for comparison. Calculated mean magnetic moment (in $\mu_B$) of Pd and Fe ions  versus impurity concentration (in at.\%) up to 25\,at.\%. }
\label{fig:Fe}
\end{figure}

The effect of iron (Fe) impurity on the Pd matrix was previously discussed in our paper~\cite{piyanzina_crystals}, where calculations were performed with lower computational precision using the VASP-5.4 version. In the current investigation, no critical concentration was found (Figure\,\ref{fig:Fe}\,a), differing from the previous results, and the general trend aligns more closely with the experimental curves~\cite{crangle,esmaeili2}, even though the measurements of Esmaeili et al. reported in~\cite{esmaeili2} were performed for thin film geometry. Furthermore, according to Ododo et al. experiments~\cite{ododo}, the critical concentration was not observed up to 0.1\,at.\%. 

The main discrepancy with the experimental trend, shown in Figure\,\ref{fig:Fe}\,a, as would be expected, appears at the ultra-low concentration region below 5\,at.$\%$, where some fluctuations in DFT values were observed. However, considering the experimental saturation magnetization around 10\,$\mu_B$~\cite{nieu}, reported in various studies, our results are just slightly higher, especially considering that our calculations are performed at 0\,K.

The calculated magnetic moment per Fe ion, the mean magnetic moment of Pd and Fe ions, and the mean of the absolute magnetic moments of Fe ions depending on impurity concentration are presented in Figure\,\ref{fig:Fe}\,b.

In Figure\,\ref{fig:Fe}\,b it is seen that Pd remains paramagnetic at 1\,at.\% of Fe, but becomes constantly polarized above this concentration, with the magnetic moment staying slightly below its bulk value. At a concentration above 13\,at.$\%$ the mean magnetic moment of Pd drops back to zero again. The magnetic moment of Fe at low impurity concentration is higher than in the bulk by $\approx$ 1\,$\mu_B$ and then gradually decays to the bulk value. 

\subsubsection{Co impurity alloy}
\label{Co_magnetism} 

\begin{figure}
\centering
\begin{minipage}{0.49\textwidth}
 \includegraphics[width=\linewidth]{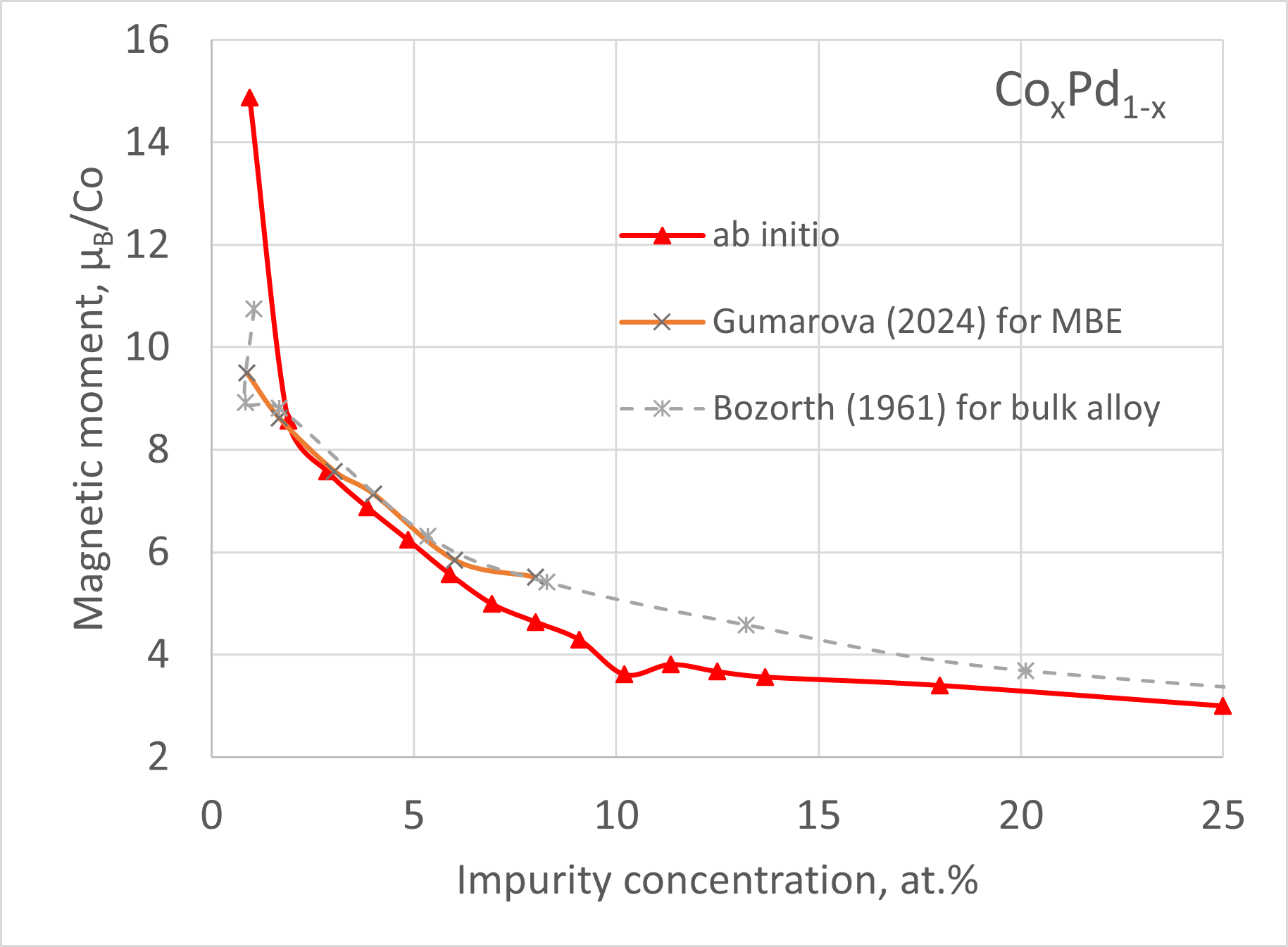} \\(a) 
\end{minipage}
\begin{minipage}{0.49\textwidth}
 \includegraphics[width=\linewidth]{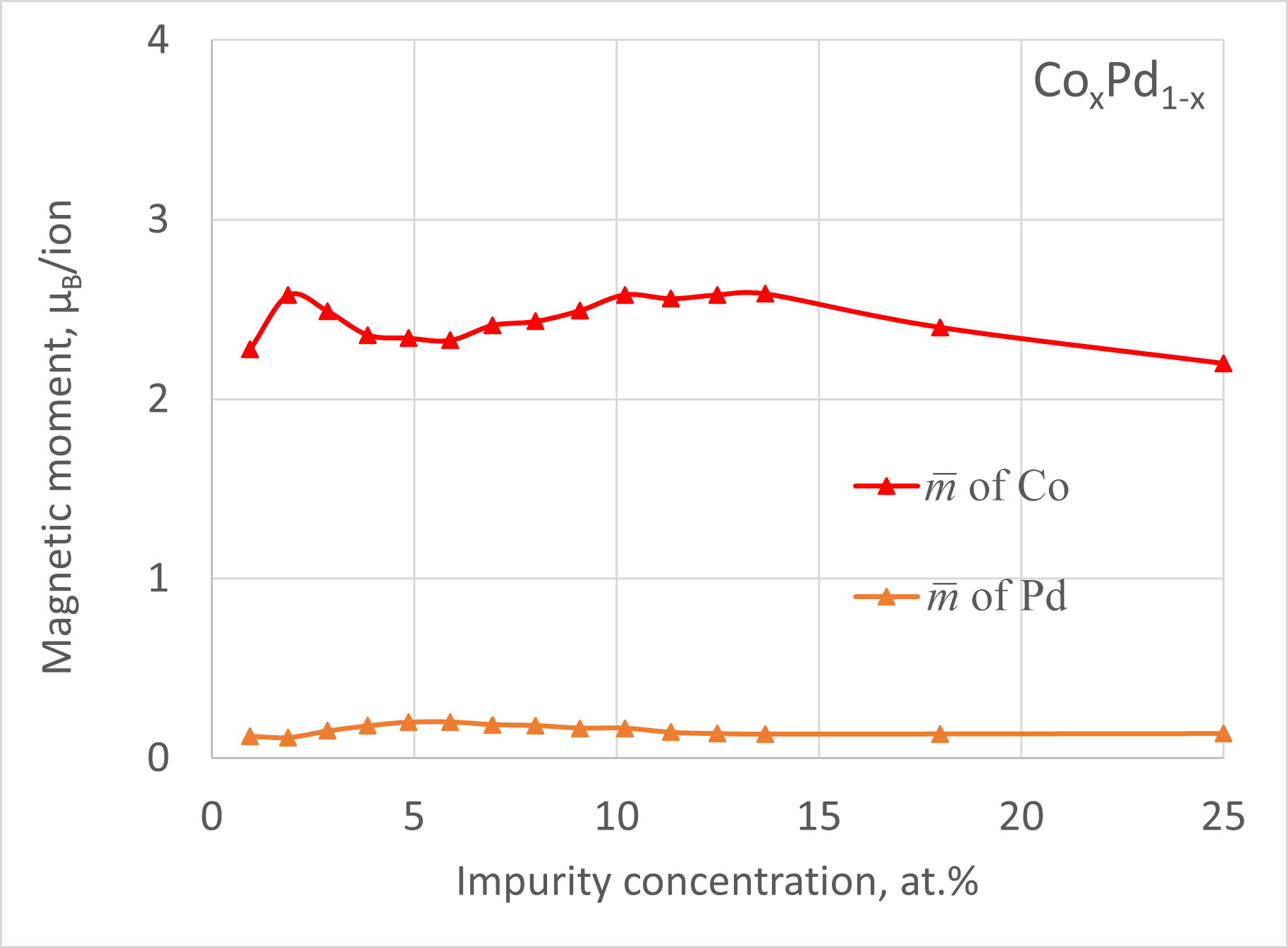} \\(b) 
\end{minipage}
\caption{(a) Calculated magnetic moment (in $\mu_B$) per impurity ion for the Co$_x$Pd$_{1-x}$ alloy versus Co concentration (in at.\%) up to 25 at.\%. The available experimental data from Refs.~\cite{gumarova,bozorth} is presented as well for comparison. (b) Calculated mean magnetic moment (in $\mu_B$) of Pd and Co ions versus impurity concentration (in at.\%) up to 25\,at.\%.}
\label{fig:Co}
\end{figure}

The cobalt (Co) impurity alloy demonstrates the most stable behavior among the alloys considered, as depicted in Figure\,\ref{fig:Co}\,a. It exhibits a strong and constant exponential decay of the magnetic moment per Co ion. The \textit{ab initio} curve in the 2-8\,at.$\%$ region aligns very well with previously published experimental data for MBE films~\cite{gumarova}, as well as bulk alloys~\cite{bozorth}. Here, similar to Fe, no critical concentration for the initialization of ferromagnetism in Pd was found for Co, suggesting that it occurs at very low concentrations. Lastly, as shown in Figure\,\ref{fig:Co}\,b, in contract to all other considered cases, Co and Pd in Co$_x$Pd$_{1-x}$ system have rather constant magnetic moment throughout the concentration range, with all impurity and matrix ions polarized in the same direction.

Older experimental measurements also revealed that no critical concentration was observed up to 0.1\,at.\% with a saturation magnetization of 10\,$\mu_B$~\cite{nieu,bozorth}. All of these findings are in reasonable agreement with our calculations.

\subsubsection{Ni impurity alloy}
\label{Ni_magnetism}

\begin{figure}
\centering
\begin{minipage}{0.49\textwidth}
 \includegraphics[width=\linewidth]{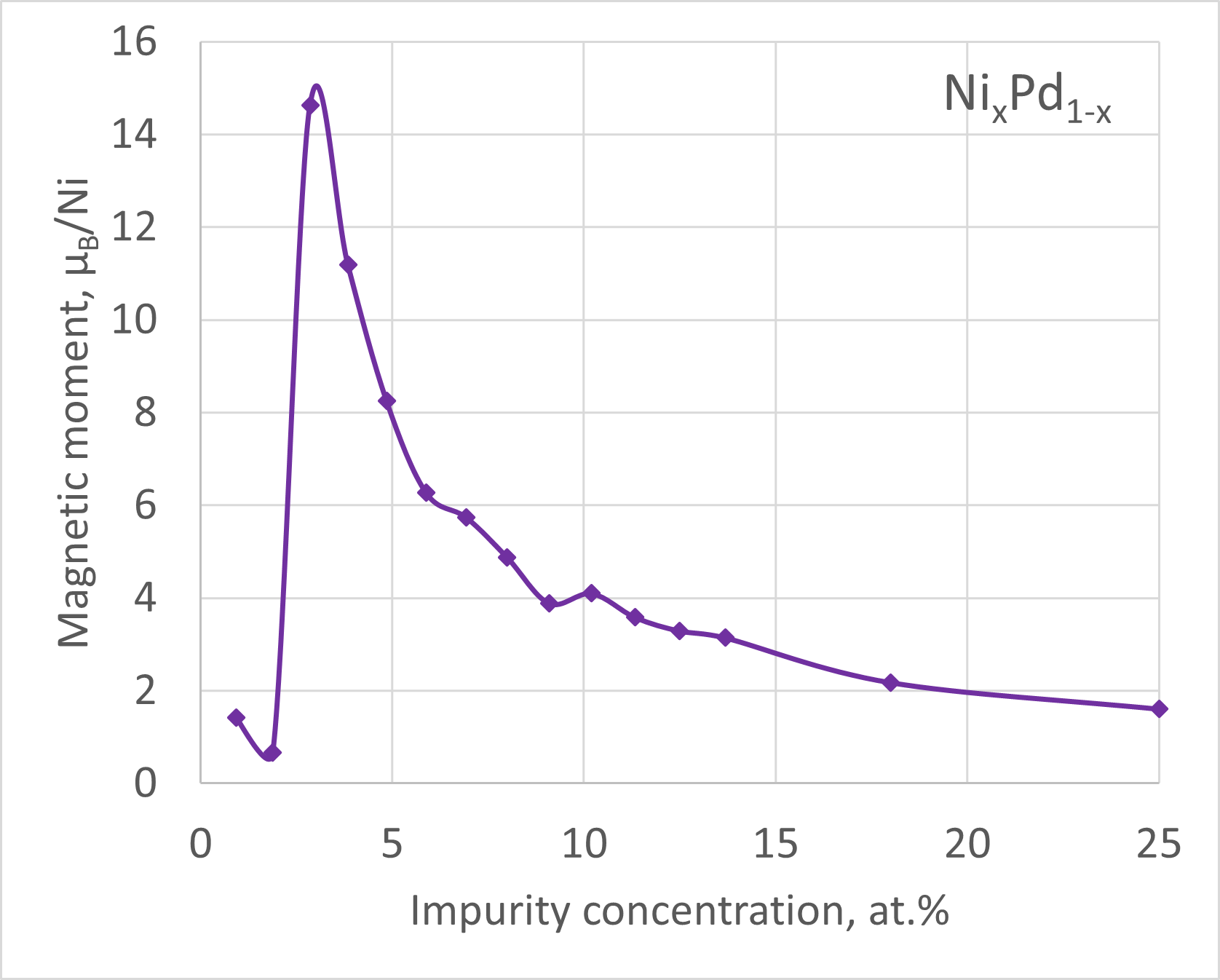} \\(a)
\end{minipage}
\begin{minipage}{0.49\textwidth}
 \includegraphics[width=\linewidth]{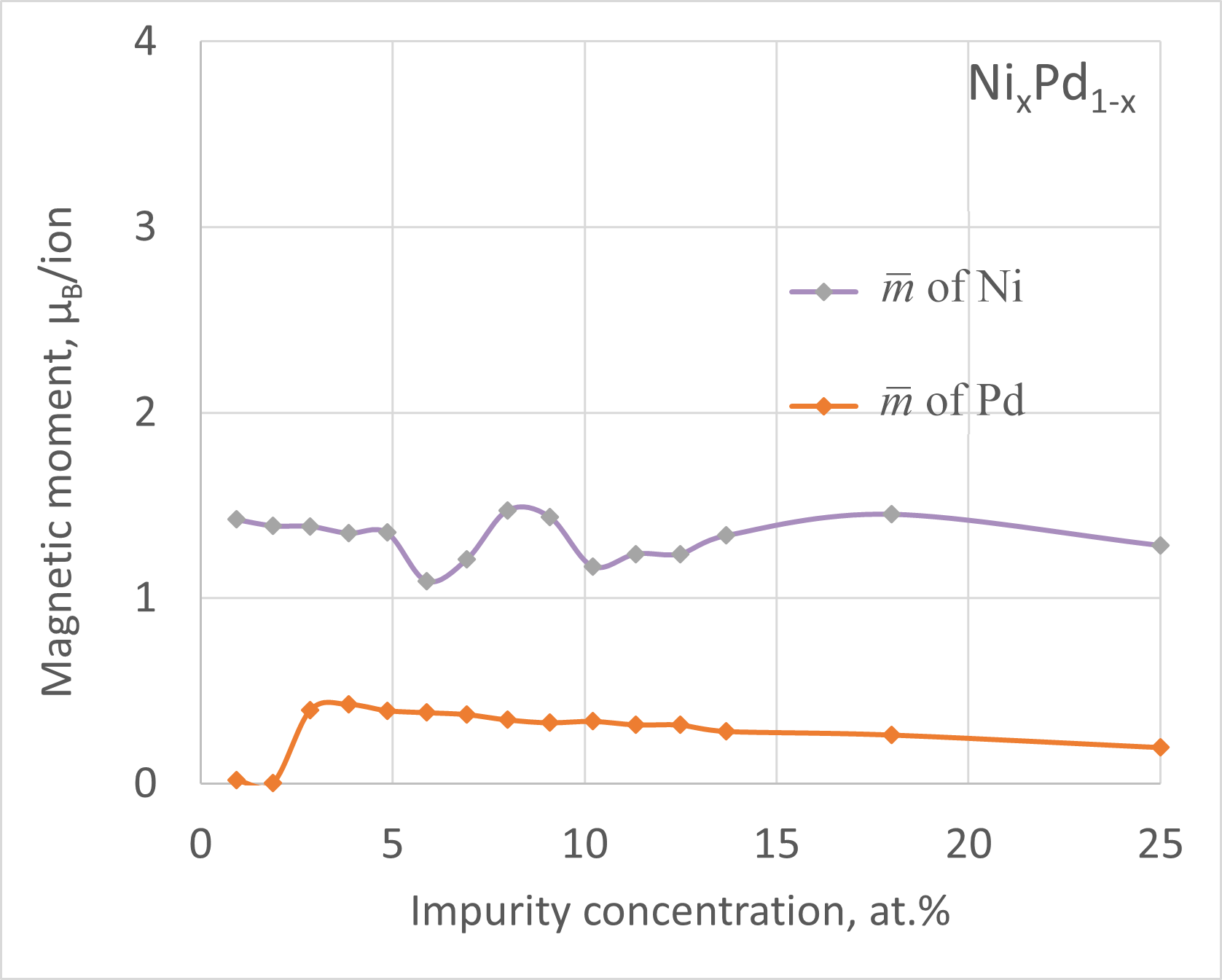} \\(b)
\end{minipage}
\caption{(a) Calculated magnetic moment (in $\mu_B$) per impurity ion for the Ni$_x$Pd$_{1-x}$ alloy versus Ni concentration (in at.\%) up to 25 at.\%. (b) Calculated mean magnetic moment (in $\mu_B$) of Pd and Ni ions versus impurity concentration (in at.\%) up to 25\,at.\%.}
\label{fig:Ni}
\end{figure}

The last alloy with nickel (Ni) impurity exhibits the presence of a clear critical concentration within the range observable by \textit{ab initio} calculations. As seen in Figure,\ref{fig:Ni} a, at 3\,at.\% the magnetic moment of Ni increases dramatically to 15,$\mu_B$ and then decreases exponentially. The concentration dependencies of the averaged magnetic moments of Pd and Ni are presented in Figure\,\ref{fig:Ni}\,b. 
Below critical concentration, Pd is paramagnetic and depolarized, but above it, Pd ions become polarized, slightly exceeding their bulk value. After reaching maximum, the mean magnetic moment of the Pd ions slowly decreases. Ni ions demonstrate some fluctuations as a result of the presence of oppositely directed magnetic moments similar to the Mn and Fe cases.

\begin{figure}
\centering
\begin{minipage}{0.49\textwidth}
 \includegraphics[width=\linewidth]{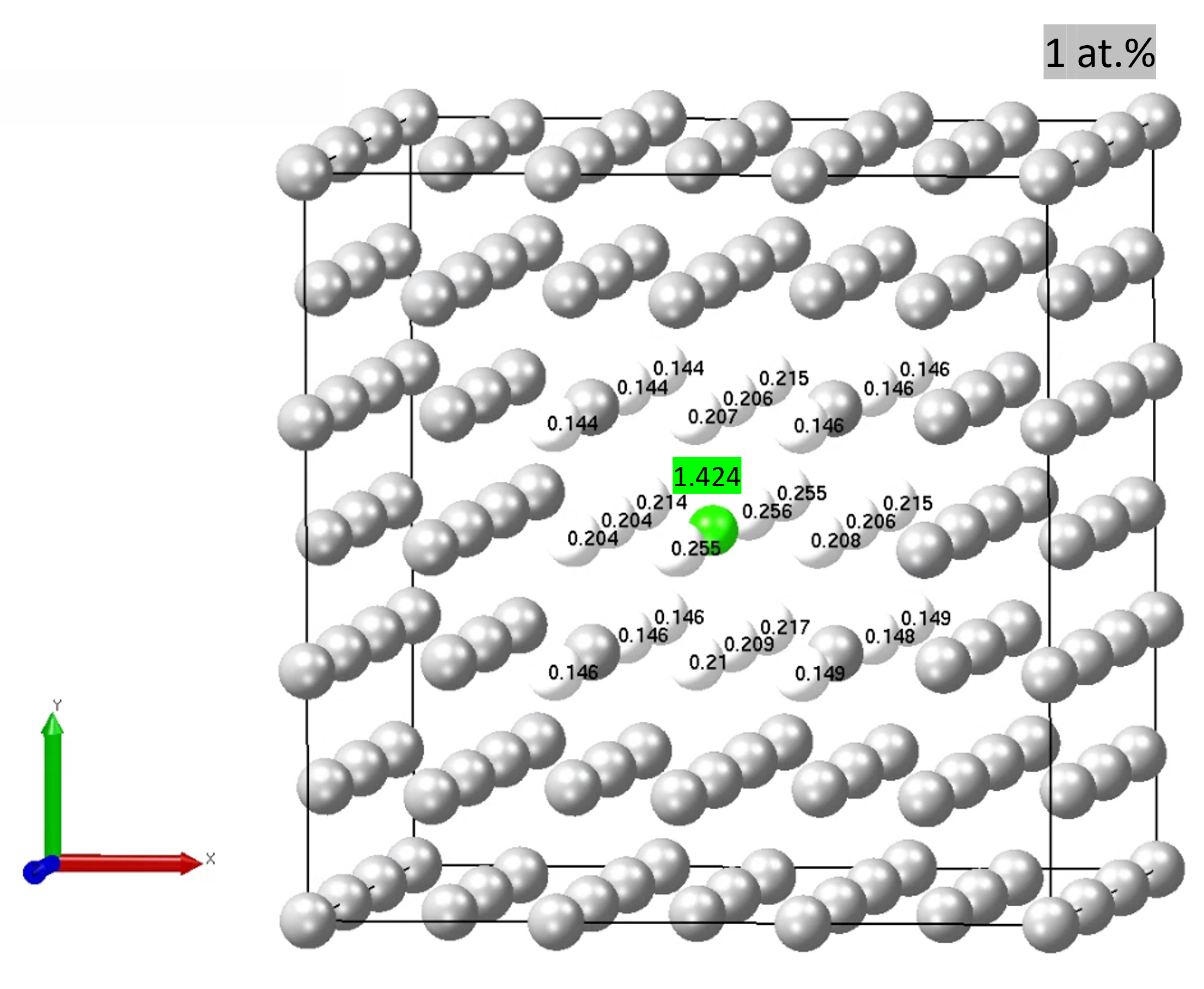} \\(a)
\end{minipage}
\begin{minipage}{0.49\textwidth}
 \includegraphics[width=\linewidth]{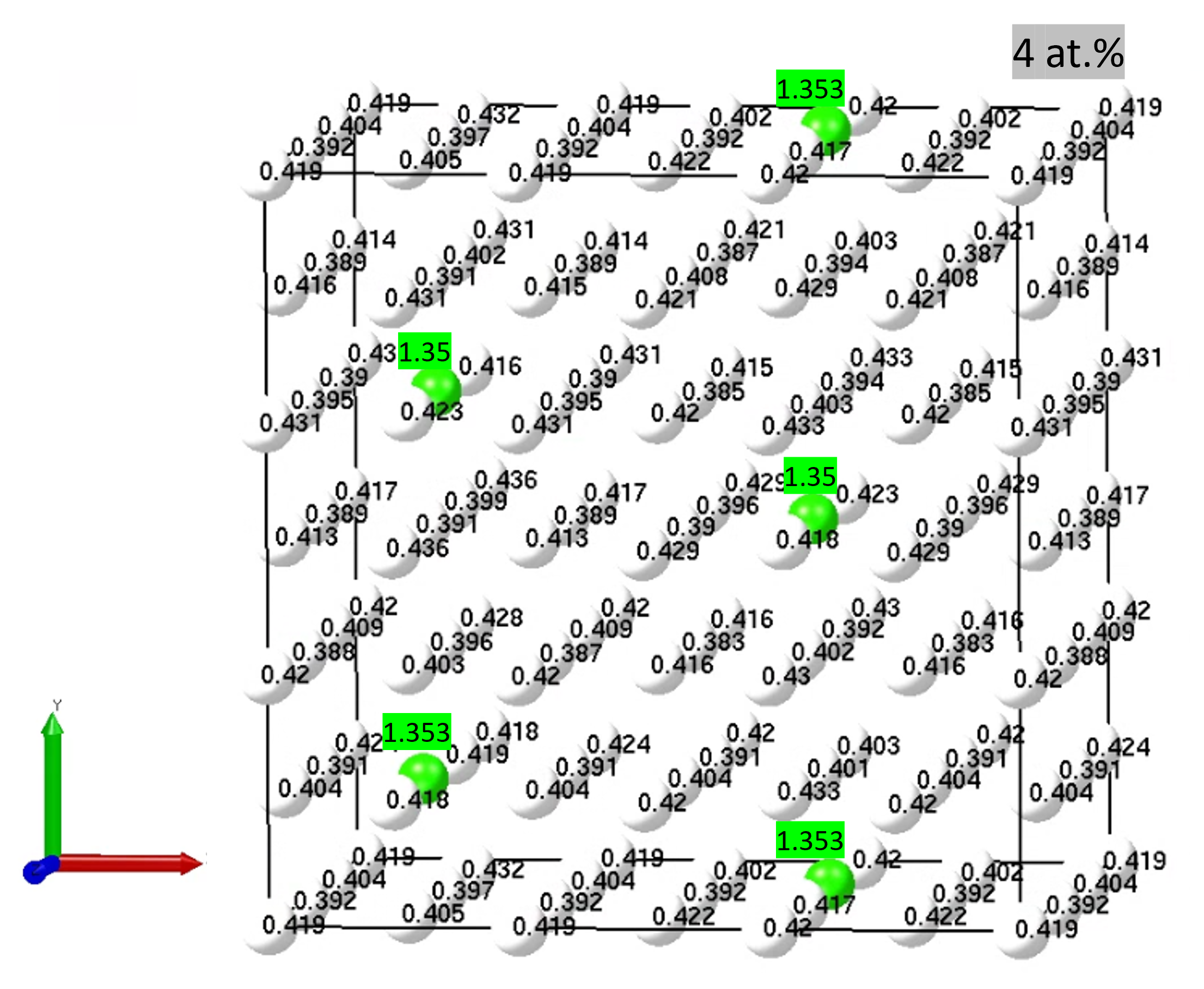} \\(b)
\end{minipage}
\caption{Distribution of magnetic moments (in $\mu_B$) in Pd$_{1-x}$Ni$_{x}$ alloys. Green spheres denote impurity Ni ions, whereas white spheres are mostly polarized Pd ions, gray spheres are Pd ions with negligible magnetization.}
\label{fig:Ni_cells}
\end{figure}

Below the critical concentration, at 1\,at.$\%$, as illustrated in Figure\,\ref{fig:Ni_cells}\,a, a single Ni ion is in the ferromagnetic state and creates a magnetic cluster around itself, with slightly polarized Pd ions (as indicated by white spheres). The other Pd ions of the matrix remain paramagnetic or oriented oppositely, resulting in a total magnetization of around 1\,$\mu_B$. As the number of Ni ions increases, their magnetic moments slightly decrease (by $\approx$ 0.1\,$\mu_B$), approaching the bulk value. At the same time, above the critical concentrations, as seen for 4\,at.$\%$ in Figure\,\ref{fig:Ni_cells}\,b, the size of the magnetic clusters is sufficiently larger, so they are able to polarize all Pd ions to $\approx$0.4\,$\mu_B$. A further increase in concentration causes these clusters to overlap, which is energetically unfavorable and consequently leads to reduced magnetization. Above $\approx$15\,at.$\%$ the magnitude of Ni's magnetic moments stabilizes, approaching the bulk value. 

The results obtained align well with the available experimental data, slightly exceeding the experimental magnetic moments for separate ions. In particular, a paper presenting Bragg scattering measurements for bulk PdNi alloys~\cite{cable} has reported magnetic moments of 1.06$\pm$0.12 for Ni and 0.14$\pm$0.02 for Pd at 8\,at.\%. Furthermore, Ododo et al. reported a critical concentration of 2.8$\pm$0.1$\%$ for NiPd alloy~\cite{ododo}, which is consistent with 3\,at.\% obtained in the present work.

\subsection{Electronic structure analysis for Ni$_x$Pd$_{1-x}$ alloy}
\label{electronic}

The observed magnetic behavior of the alloys can be explained by analyzing the distribution of their electronic states. Density-of-state (DOS) plots provide valuable insight into the emergence of magnetism. Since the Ni$_x$Pd$_{1-x}$ alloy exhibit the most pronounced transition from paramagnetic to ferromagnetic state at critical concentration of 3\,at.$\%$, we calculated atom-, spin- and orbital-resolved DOS plots both above and below this concentration, as well as for higher concentrations to follow the evolution of the electronic and magnetic states. 

\begin{figure}[ht]
\centering
\includegraphics[width=8 cm]{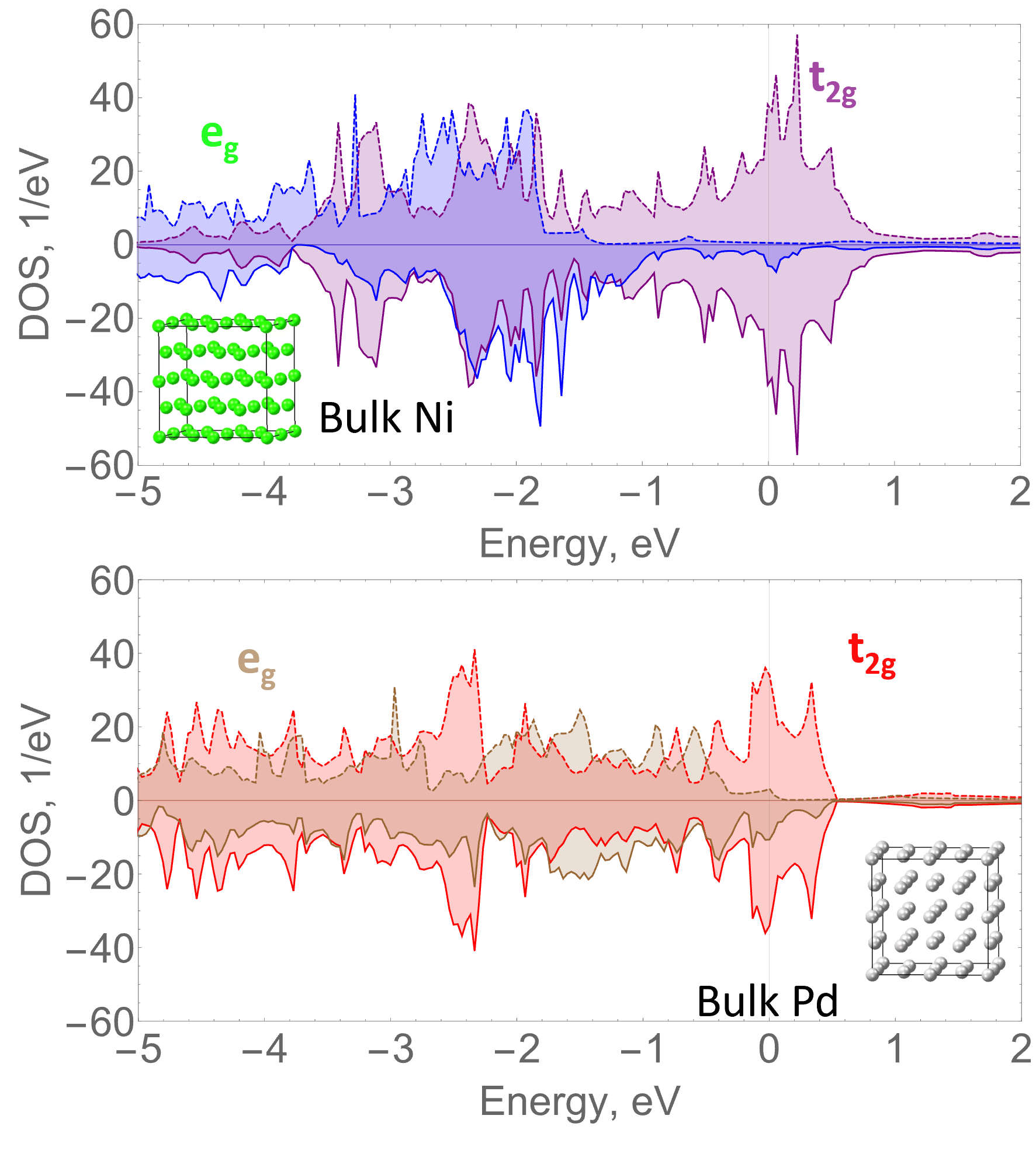}
\caption{The calculated DOS for bulk $2\times2\times2$ cell of fcc Ni and fcc Pd with orbital resolution. That is, t$_{2g}$ and e$_g$ orbitals are presented for spin up and down components. The corresponding unit cells are illustrated as a caption.} 
\label{fig:Ni_dos_orbitals}
\end{figure}

For comparison, we first present the DOS for bulk Ni and Pd (2$\times$2$\times$2 cells) with orbital resolution, highlighting the contributions of the orbitals t$_{2g}$ and e$_g$ (Figure\,\ref{fig:Ni_dos_orbitals} with the computational cells shown in the caption). It is well known that Ni is a good conductor of electricity, which is reflected in the significant t$_{2g}$ contribution at the Fermi level and the presence of the states across the entire energy range, both below and above the Fermi energy. The energy splitting responsible for the magnetic behavior is mainly observed in the e$_g$ states, which are located mostly below the Fermi level, with only the spin-down component crossing the Fermi level. This splitting is crucial in driving the ferromagnetic character of Ni. 

As seen for the DOS of the bulk Pd (Figure\,\ref{fig:Ni_dos_orbitals}), the main contribution to conductivity is also from electrons of the t$_{2g}$ orbitals, whereas the splitting occurs in e$_g$ orbitals energy levels. This difference in spin up and down occupancy is sufficiently smaller than in the bulk Ni case, which consequently results in a smaller bulk magnetization, as was mentioned in Section\,\ref{bulk} (Table\,\ref{tab:bulk}).

\begin{figure}
\centering
\includegraphics[width=13 cm]{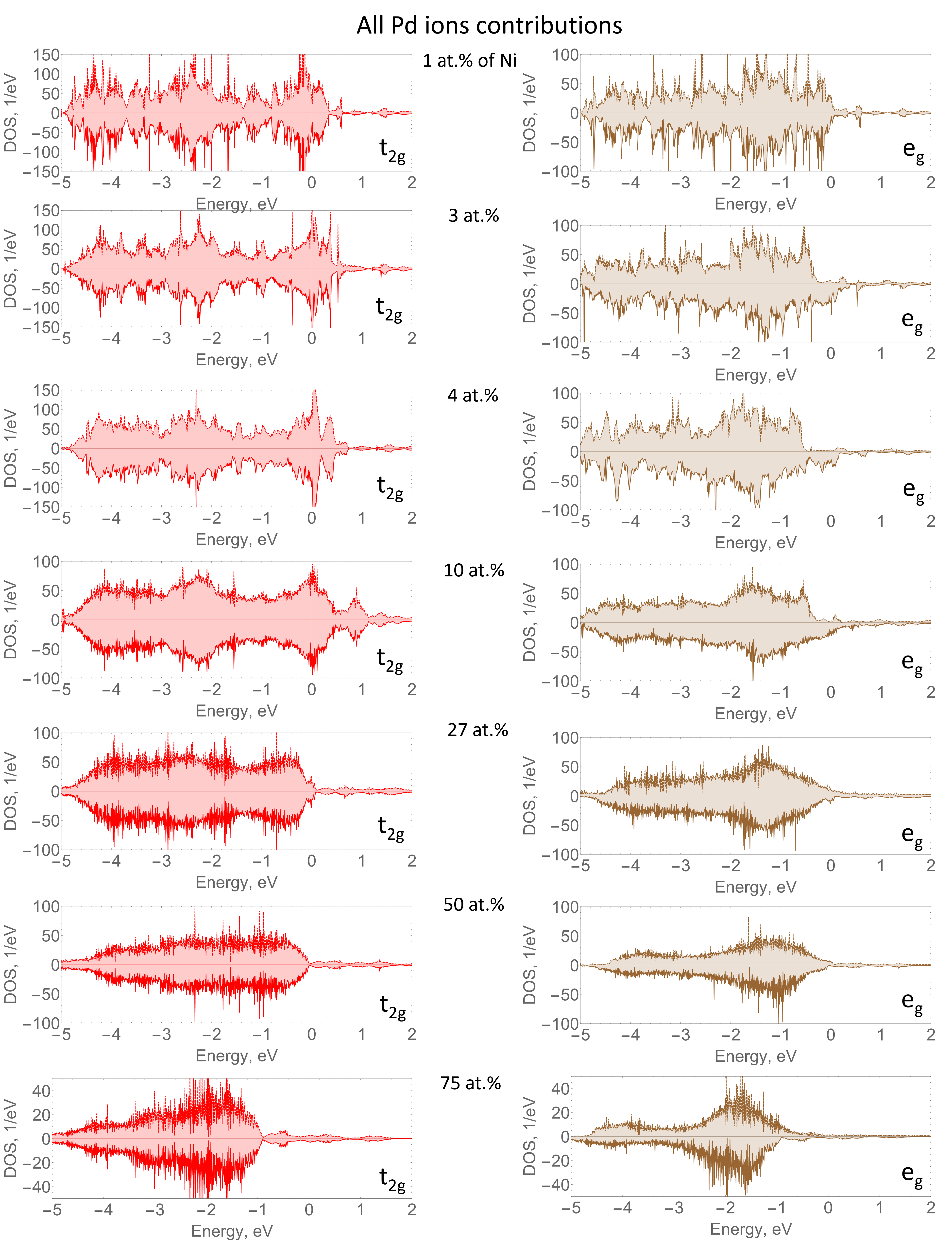} 
\caption{The calculated orbital-resolved DOS for Pd ions of Ni$_x$Pd$_{1-x}$ alloy with denoted concentrations. The spin up and down components for the orbitals t$_{2g}$ and e$_g$ are presented.} 
\label{fig:NiPd_dos_orbitals_Pd}
\end{figure}

In the next step, the DOS spectra for the PdNi alloy with various concentrations of Ni were calculated, up to 75\,at.\%. Figure\,\ref{fig:NiPd_dos_orbitals_Pd} illustrates the orbital-resolved DOS plots for Pd ions. Specifically, the figure shows the orbitals t$_{2g}$ and e$_g$ as the sum of all contributions of Pd ions. It is seen that at low Ni concentrations (Pd-rich alloys, below the critical concentration), the states t$_{2g}$ cross the Fermi level, contributing predominantly to conductivity, while the density of e$_g$ states at the Fermi level is much lower. The overall distribution of spin components (up and down) is symmetric, indicating paramagnetic behavior. 

Above the critical concentration of Ni impurity, while the t$_{2g}$ states remain relatively unchanged, the distribution of the e$_g$ orbitals undergoes a significant transformation. The carriers with spin-up components flip their spins. This leads to an increase in the magnetic moment of Pd ions, which is consistent with the maximum observed in Figure\,\ref{fig:Ni}\,a. Consequently, the contribution to the conductivity of electrons in the e$_g$ states increases.

A further increase in Ni concentration (as presented for 4 and 10\,at.\%) results in a reduction in the density of the e$_g$ states at the Fermi level and a downward shift in the t$_{2g}$ states. At concentrations greater than 27\,at.\%, both t$_{2g}$ and e$_{g}$ states are predominantly located below the Fermi level, with a more homogeneous distribution of spin projections (up and down), leading to a decrease in overall magnetization. The downshift of the Pd \textit{d} states significantly suppresses the conductivity of the Pd ions.

\begin{figure}
\centering
\includegraphics[width=13 cm]{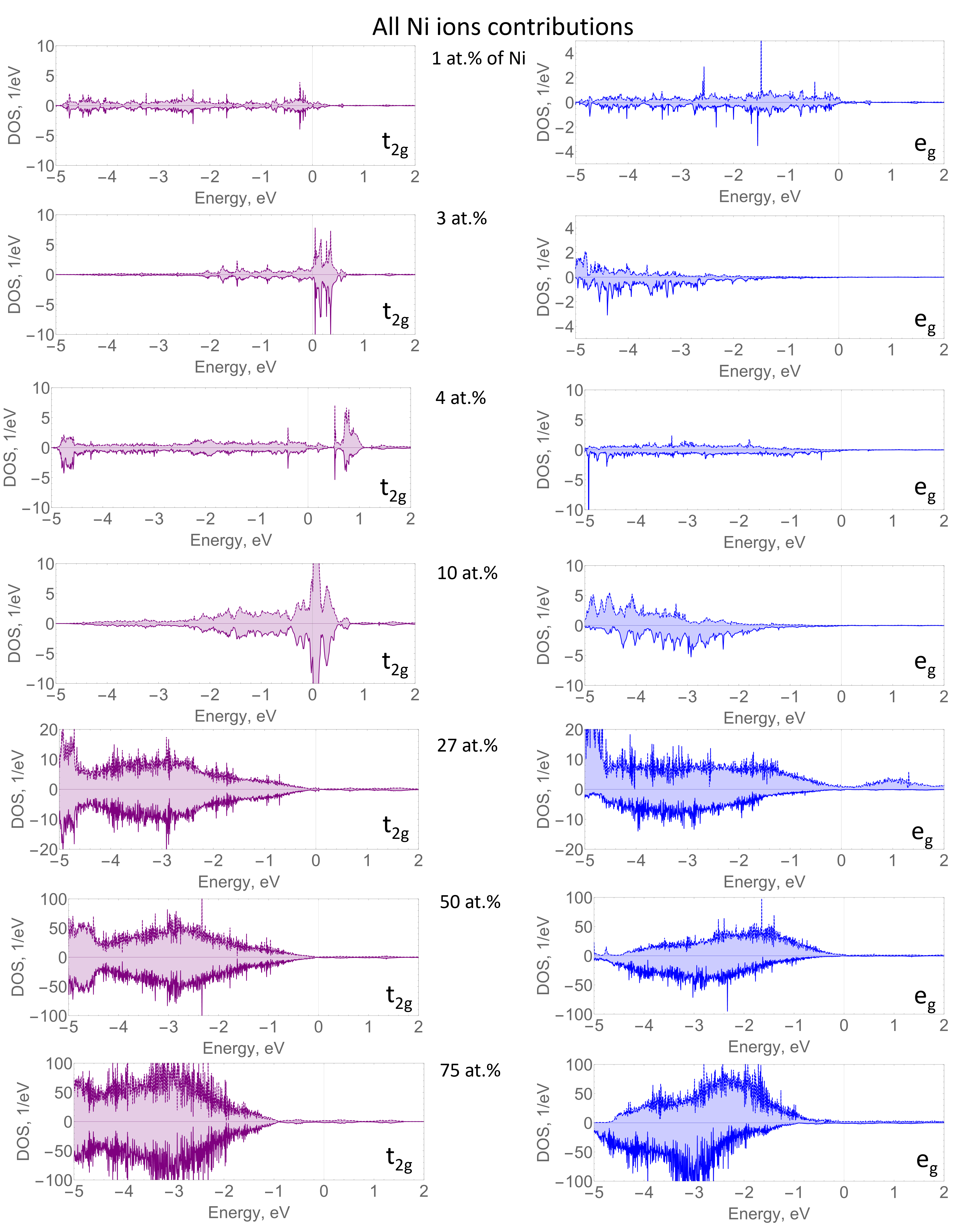} 
\caption{The calculated orbital-resolved DOS for Ni ions of Ni$_x$Pd$_{1-x}$ alloy with denoted concentrations. The spin up and down components for the orbitals t$_{2g}$ and e$_g$ are presented.} 
\label{fig:NiPd_dos_orbitals_Ni}
\end{figure}

The contribution of Ni ions is presented in separate plots, as shown in Figure\,\ref{fig:NiPd_dos_orbitals_Ni}. At ultra-low Ni concentrations, below the critical concentration, Ni ions exhibit negligible magnetization, with orbital states t$_{2g}$ and e$_{g}$ located primarily in the valence band. However, at concentrations above the critical point (3\,at.$\%$), a noticeable up-shift is observed in the t$_{2g}$ states and a down-shift is observed in the e$_{g}$ states. At 10\,at.$\%$ the patterns of states and their energy positions for Ni closely resemble those of the bulk Ni, as shown in Figure\,\ref{fig:Ni_dos_orbitals}. In particular, the t$_{2g}$ states display a high DOS near the Fermi level, while the spin-up component of the e$_{g}$ states is sufficiently suppressed.

As the Ni content increases, the states shift downward in the energy scale, leading to a reduction in conductivity through the Ni t$_{2g}$ states, similar to the behavior observed for Pd. The energy splitting in the e$_{g}$ states, responsible for Ni magnetization, also changes, with the spin-up component becoming more dominant. That tendency was also observed for Pd. In general, the splitting of the e$_g$ orbitals and the resulting giant magnetization of Ni and Pd ions at concentrations between 3 and 10\,at.$\%$ exhibit behavior similar to that of bulk Ni. 

Pd and Ni are well-known good conductors, as reflected in Figure\,\ref{fig:Ni_dos_orbitals}, where both elements show substantial contributions at the Fermi level. In low-Ni regime alloys, conductivity is primarily dominated by Pd ions. However, at a concentration of 50\,at.\% both Pd and Ni ions have a similar magnitude of DOS at the Fermi level, indicating a shared role in the conductivity of the alloy. 

As the Ni content further increases, the DOS at the Fermi level decreases. This behavior can be attributed to the following factors:
at low concentrations of impurities above the critical point, all \textit{d} states are located near the Fermi level and can contribute to conductivity. However, at higher concentrations, particularly around 27\,at.\%, these states become filled as the number of free valence electrons of Ni having similar energy increases, reducing the contribution of both Pd and Ni at the Fermi level, as shown in Figures,\ref{fig:NiPd_dos_orbitals_Pd}, \ref{fig:NiPd_dos_orbitals_Ni}. At 50\,at.\% Ni, most states are filled, resulting in available states only in the valence band. The situation remains unchanged at 75\,at.\% Ni.

\section{Discussion}

According to the theory, magnetization arises from the indirect interaction of impurity spins mediated by the strongly correlated \textit{d}-electrons in the Pd matrix. Each impurity ion creates a cloud of polarized electrons, and once these clouds overlap, the indirect exchange interaction contributes to ferromagnetism~\cite{nieu}. The strength of this contribution exceeds that of the RKKI (Ruderman, Kittel, Kasuya, and Yosida) interaction~\cite{rkki} due to the relatively small distances between impurities. However, below the critical concentration, the RKKI contribution dominates, resulting in spin-glass behavior. This state is characterized by the randomness of spin orientations and the lack of long-range magnetic order.

From a band structure perspective, below the critical point, the spin states of both Pd and the impurity ion exhibit a more homogeneous distribution and filling, indicative of delocalized electronic behavior. This delocalization prevents the establishment of the ferromagnetic order. Among the alloys considered, the condition for the ferromagnetic contribution exceeding the RKKI is obviously different because of different magnitude of exchange interactions.

Slightly above the critical concentration, ferromagnetic ordering emerges as the impurity concentration increases, and the overlap between polarized electron clouds becomes sufficient to enhance the ferromagnetic exchange interaction. This results in the alignment of the spins. The rise in ferromagnetic behavior can be traced back to changes in the DOS, particularly the population of spin-up states.

At higher concentrations, the decline in magnetization is attributed to the progressive filling of vacant spin-down states. As more states below the Fermi level become occupied by spin-down electrons, the overall magnetic moment decreases. This filling effect reduces the imbalance between the spin-up and spin-down states, leading to a reduction in net magnetization.   

The reasoning presented is true for all cases, with some differences coming from the different filling of \textit{d} orbitals with electrons. The summary of the characteristics obtained by DFT for the alloys analyzed is collected in Table\,\ref{tab:final}. Overall, the magnitude of maximal magnetic moment was found to be similar for all alloys, but the critical concentrations varies significantly without any obvious pattern. Regarding the critical concentration values, they are approximately 3\,at.\% for both MnPd and NiPd alloys. For FePd and CoPd, the critical concentration is estimated to be below 1\,at.\%, since our computational capabilities are limited to explore even lower concentrations. Moreover, the alloy with manganese impurity ions displays a second critical concentration, when the second maximum is observed, which could be related to the intrinsic antiferromagnetic nature of manganese.    
\begin{table}[ht]
    \centering
    \caption{Summary of characteristics obtained for Pd$_{1-x}$$M_{x}$ alloys ($M$ =  Mn, Fe, Co, Ni) by DFT calculations: maximal magnetic moments per impurity content (\textit{m}), critical concentrations if observed (\textit{c}), and magnetic ordering in alloys (fM is for ferrimagnetic, FM is for ferromagnetic, respectively).}
    \label{tab:final}
    \begin{tabular}{l|llll}
         Alloy & MnPd & FePd & CoPd & NiPd \\ \hline
      \textit{m}, $\mu_B$    & 13 & 13 & 15 & 15  \\ 
       \textit{c}, at.\,\%  & 3 & - & - & 3  \\  
      magnetic ordering  &  fM &  fM & FM  & fM  
    \end{tabular}
    \end{table}

Lastly, all alloys with magnetic impurities, except those with Co, exhibit rather ferromagnetic behavior at least down to certain concentrations. Complete alignment of the magnetic moments occurs at concentrations greater than approximately 15\,at.\% when fluctuations in the magnitude of the total magnetic moment, as well as those of impurities and Pd ions, stops (Figures\,\ref{fig:mn}, \ref{fig:Fe},  \ref{fig:Ni}).

\section{Conclusion}
\label{concl}

In the present work, we have demonstrated that calculations within density functional theory are able to reproduce the main experimental finding of alloys of Pd$_{1-x}$$M_x$ ($M$ =  Mn, Fe, Co, Ni) in most of the concentration range (0.01$<$x$<$1). For all alloys considered, the maximum magnetic moment per impurity content varies in the range of 13-15\,$\mu_B$. This value is several times higher than the magnetic moments of the bulk components of the alloys. The critical concentration for PdNi and PdMn was found to be equal to 3\,at.\%, which is consistent with the experimental data available for NiPd. Moreover, all of the alloys demonstrated a decrease in the magnetic moments with increasing impurity concentration, reaching the value of the bulk component at the rich-impurity regime. However, the character of the curves differs for different alloys. 
The reason of observed differences are not unambiguous and will be the subject of further research as well as the magnetic anisotropy features which may be observed in these systems.

\section*{Acknowledgments}
Computational resources were provided by the Laboratory for the computer design of new materials and machine learning of Kazan Federal University in cooperation with the Atomdata-Innopolis data center. 

R.Burganova, A.Gumarov and I.Yanilkin acknowledge the Kazan Federal University Strategic Academic Leadership Program (PRIORITY-2030).
The work of I.Piyanzina was supported by the grant 24PostDoc/2-2F006.
V.Stolyarov thanks the grant of the Ministry of Science and Higher Education of the Russian Federation, Grant No. 075-15-2024-632 for supporting the electronic structure calculations.

\end{document}